\documentclass[conference]{IEEEtran}
\IEEEoverridecommandlockouts
\usepackage{cite}
\usepackage{amsmath,amssymb,amsfonts}
\usepackage{algorithmic}
\usepackage{float}
\usepackage{graphicx}
\usepackage{mdframed}
\usepackage{textcomp}
\usepackage{booktabs}
\usepackage{tcolorbox}
\usepackage{listings}
\usepackage{xcolor}
\usepackage{svg}
\usepackage{pgfplots}
\pgfplotsset{width=10cm,compat=1.9}
\usepackage{array}
\usepackage[nolist,nohyperlinks]{acronym}
\usepackage[normalem]{ulem}
\def\BibTeX{{\rm B\kern-.05em{\sc i\kern-.025em b}\kern-.08em
    T\kern-.1667em\lower.7ex\hbox{E}\kern-.125emX}}

\definecolor{nicolasred}{rgb}{0.7804, 0, 0.0431}
\definecolor{nicolascyan}{rgb}{0.1882, 0.7137, 0.7725}

\definecolor{ali}{rgb}{0.9, 0.1, 0.9}
\definecolor{done}{rgb}{0.298, 0.733, 0.090}

\tcbuselibrary{listings, skins} 
\definecolor{promptbackground}{rgb}{0.95, 0.95, 0.95}

\begin{document}

\title{\huge{Telco-RAG: Navigating the Challenges of Retrieval Augmented Language Models for Telecommunications}}

\author{
 Andrei-Laurentiu Bornea$^*$,
 Fadhel Ayed$^*$,
 Antonio De Domenico$^*$,
 Nicola Piovesan$^*$, 
 Ali Maatouk$^+$ \\
$^*$Paris Research Center, Huawei Technologies, Boulogne-Billancourt, France\\
$^+$Yale University, New Haven, Connecticut, USA 
\vspace{-1em}}

\maketitle

% {\color{done}[Ali]: I did changes to the abstract's structure, just double check if the meaning you intended is still preservered}
\begin{abstract}
The application of Large Language Models (LLMs) and Retrieval-Augmented Generation (RAG) systems in the telecommunication domain presents unique challenges, primarily due to the complex nature of telecom standard documents and the rapid evolution of the field. 
%Our study advances state-of-the-art retrieval-augmented models for the telecommunication domain by optimizing chunk sizes, embedding models, indexing strategies, incorporating conversational prompts, and query refinement. This study highlights that limited chunk sizes significantly improve accuracy in processing technical standards.
The paper introduces Telco-RAG,\footnote{https://github.com/netop-team/telco-rag} an open-source RAG framework designed to handle the specific needs of telecommunications standards, particularly 3rd Generation Partnership Project (3GPP) documents. 
%It is engineered to improve how professionals access and adhere to these standards, thus enhancing development cycles and regulatory compliance.
Telco-RAG addresses the critical challenges of implementing a RAG pipeline on highly technical content, paving the way for applying LLMs in telecommunications and offering guidelines for RAG implementation in other technical domains.

\end{abstract}

% {\color{done}[Ali]: precision retrieval sounds off}
% {\color{done}[Ali]: lets use a different word than markedly}

% \begin{IEEEkeywords}
% component, formatting, style, styling, insert
% \end{IEEEkeywords}

\section{Introduction}
% \textcolor{green}{Antonio: we need 2 sentence explaining why telecom needs LLM}

% {\color{done}[Ali]: The introduction is generally good. What needs to be improved in it is the information flow. Try to ensure that each paragraph has a unique message to give, and with a smooth transition to the next message. }

Large language models (LLMs) are designed to understand, generate, and process text by leveraging extensive training data. These models, built upon architectures such as Transformers, employ deep learning techniques to analyze and predict language patterns\cite{vaswani2017attention}.
%\sout{LLMs, which are based on transformer architectures, have made significant progress in the field of natural language processing. They are particularly effective in various applications including translation, summarization, and conversational interactions, thanks to their powerful computational abilities and extensive training on large datasets.} 
Their capabilities are largely attributed to the vast amount of text they process during training, allowing LLMs to develop a nuanced understanding of language, context, and even idiomatic expressions. The utility of LLMs extends across various domains, among which telecommunications, where models can improve operational efficiency and enhance customer satisfaction \cite{maatouk2024large}.

 % {\color{done}[Ali]: reduce a bit the claims}
 % {\color{done}[Ali]: The transition from LLMs in general to RAG can be improved. Perhaps instead of discussing general problems with LLMs, we can say that training them on domain data is very demanding and then in the next paragraph, also hallucinations and grounding answers in documents, RAG comes as a solution}

 Standalone language models rely solely on their internal representations and learned parameters to generate text, which showcases modest knowledge in technical domains such as telecommunication standard documents \cite{maatouk2023teleqna}. 
 % \textcolor{blue}{Antonio: better cite https://arxiv.org/abs/2401.08406 or this https://arxiv.org/abs/2312.05934 Avoid website and blog}
 Two primary methodologies have emerged to address this challenge: fine-tuning and retrieval-augmented generation (RAG). Fine-tuning enables language model specialization via further training of a fraction of the parameters using a domain-specific dataset. However, fine-tuning can incur a high computational cost \cite{thompson2020computational} and is not suited for rapidly evolving domains where new knowledge needs to be incorporated on a regular basis. 
 RAG stands out as an appealing alternative due to its cost-effectiveness, adaptability, and scalability \cite{ovadia2024fine}. In the RAG paradigm, knowledge from external sources is fetched in real-time when a query is addressed to the system. This is particularly tailored for quickly evolving fields \cite{balaguer2024rag}. 
 %Moreover, RAG offers greater adaptability and scalability compared to fine-tuned models, particularly in their application across multiple specific domains,\cite{siriwardhana2023improving. 

% {\color{ali}[Ali]: Reference needed. Perhaps Teleqna?}
% {\color{done}[Ali]: The paragraph is very hard to follow. We talk about studies to optimize RAGs and we do not mention what they do. It would be better to talk about what people do, and how 3GPP stuff are different, and which brings us to the stuff}
% {\color{done}[Ali]:Break the following into two sentences; always privilege smaller sentences}

% {\color{ali}[Ali]: This paragraph is not useful as it repeats last paragraph idea, i say delete it so to gain space} 
% Recent advancements in RAG systems have sparked significant interest, particularly in their application across multiple specific domains \cite{siriwardhana2023improving}. RAG systems enhance the capabilities of Large Language Models by integrating external knowledge sources, thereby improving accuracy and relevance in responses through context-specific retrieval.

% {\color{ali}[Ali]: Write a sentence first about why the interest in RAG systems here such as chatbots for professionals, etc. Also, the last sentence we don't really care about as its too general (we want to focus on telecom stuff). Aim to avoid saying things like we want to do it because its a good test. A research problem has to have value not just because it is an interesting problem to work on.}

In the telecommunication industry, a retrieval-augmented language model that masters complex industry-specific knowledge, such as the content of technical standards, would hold significant practical value \cite{piovesan2024telecom}. For instance, it would allow the development of an advanced chatbot for professionals. Such a tool would increase the accuracy and speed with which telecommunications professionals access and comply with international standards, fostering quicker development cycles and improved regulatory adherence. 

% {\color{done}[Ali]: I feel we are emphasizing on RAG systems as a whole rather than 3GPP. The readers are going to be telecom-oriented rather than general machine-learning people. Therefore, instead of positioning the study as a test for RAG systems etc., let us say that it solves the problem of dealing with standards documents}

% {\color{ali}[Ali]: Yes, this paragraph is good. Emphasize on this and try to reduce the size of the previous two paragraphs. }

% {\color{done}[Ali]:Once you define an acronym, no need to keep redefining it}
% {\color{done}[Ali]:Same comment as the above, it is way too general about RAG systems. The first paragraph is general and has been talked about before so it doesn't add much to the message (I suggest removing it)}
% \textcolor{cyan}{[Nicola: I agree with Ali. I would remove completely the previous paragraph.]}
% {\color{ali}[Ali]: It feels like we are repeating ourselves and for 4 paragraphs talked about RAG being important in telecom and so on. Make things more concise, simplified and straight to the point. Why we need to have good standards comprehension (applications), then challenges and what others have done, and then you can state the contribution. This can gain you a lot of space for later. }

In this work, we concentrated our efforts on telecommunication standards, and specifically 3rd Generation Partnership Project (3GPP) documents. This focus was motivated by the aforementioned practical utility of a chatbot specialized in 3GPP and by the observation that even state-of-the-art language models, such as GPT-4, exhibit scarce knowledge of this content \cite{maatouk2023teleqna}. 

We have identified that the conventional RAG setup, which typically extracts three to five data segments of 512 tokens each \cite{llamaindex2024}, does not adequately meet the intricate demands of telecommunications standards. Consequently, we have developed a specialized RAG pipeline named Telco-RAG specifically optimized for 3GPP documents. 

Besides, through our design and methodology, we aim to provide generally applicable guidelines to overcome the common challenges faced when implementing an RAG pipeline in highly technical domains. These include identifying the most impactful hyperparameters to tune, recommending default settings \cite{finardi2024chronicles}, reducing the high random access memory (RAM) usage, and refining the user's query \cite{chan2024rqrag}. We
expect that the Telco-RAG, which we make publicly
available as an open-source chatbot for 3GPP standards, and the associated results will contribute substantially to integrating AI in the telecommunications field.

\section{Methodology}\label{sec:methodology}
% \textcolor{green}{Antonio: start presenting briefly how RAG works and the figure of the overall architecture, which you need to elaborate more showing the relations between the different blocks and the associated input and output}

% \textcolor{blue}{Antonio: in this section, every single block of the architecture has to be presented.}
% \textcolor{blue}{Paragraph introducing the general core components of RAG (chunking, embedding, similarity search).}
\begin{figure*}[ht]
\centering
\includegraphics[width=1\textwidth]{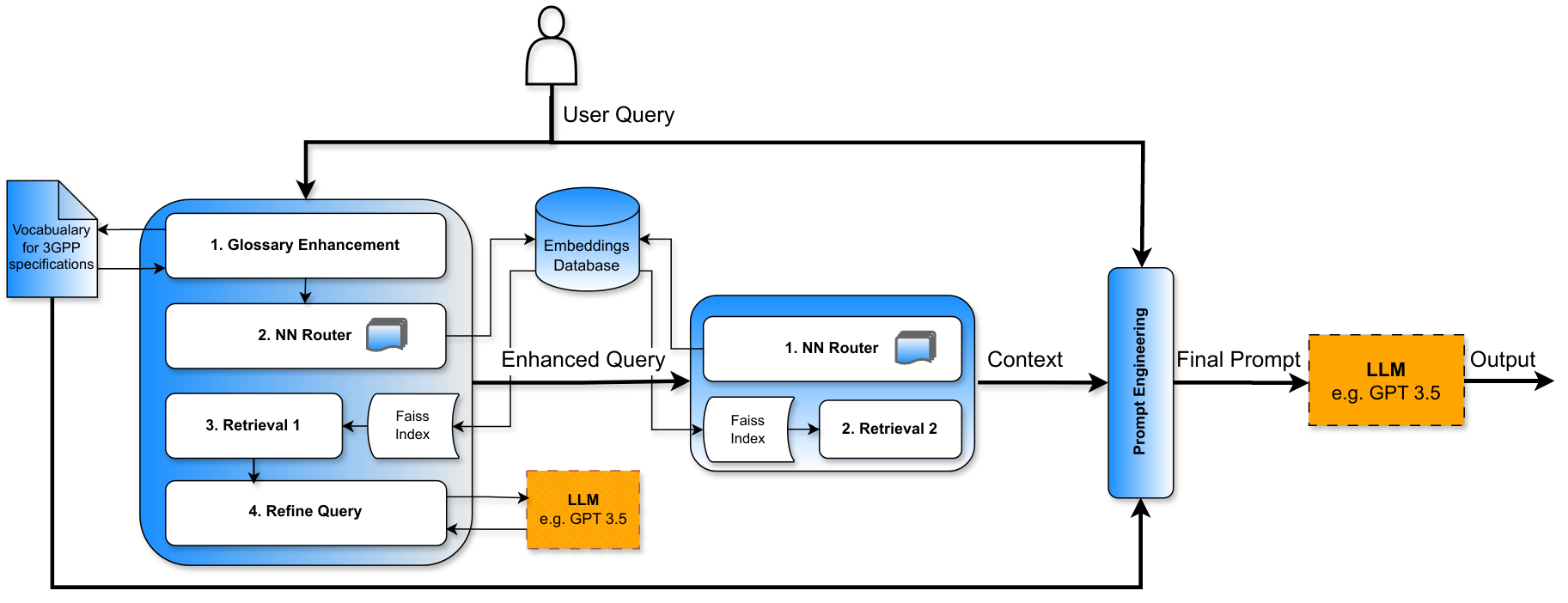}
\caption{The proposed Telco-RAG architecture.}
\label{fig: pipeline}
\end{figure*}

RAGs improve the quality of the LLM-generated responses by providing the LLM with external sources of knowledge based on a large corpus of documents.
RAG pipelines start by splitting the document corpora into fixed-sized long (chunk size) segments called chunks. Using an embedding model, each chunk is transformed into a vectorial representation capturing the semantics of the segment. When a query is presented, the system identifies the relevant chunks by computing a similarity between the chunks' embeddings and the query's embedding. Lastly, RAG presents the relevant chunks, called the context, alongside the query to a LLM that generates the final response.

Any implementation of a RAG system for telecommunications will face four critical challenges: sensitivity to hyperparameters \cite{finardi2024chronicles}, vague user queries \cite{chan2024rqrag}, high RAM requirements, and sensitivity to the quality of the prompts \cite{unleashing2023prompt}. In particular, poor prompts affect the capacity of the LLMs to comprehend the context of the queries and correctly reply to them. Moreover, vague queries limit the precision of the retrieval stage.

Fig.~\ref{fig: pipeline} depicts the proposed Telco-RAG tailored for LLM deployment in the telecommunications sector. The proposed system aims to improve the retrieval and processing of technical documents, focusing specifically on 3GPP standards. It features a dual-stage pipeline, including a query enhancement stage and a retrieval stage. The query enhancement stage includes four steps: initially, it employs a custom glossary of technical terms to augment the query, enhancing contextual understanding. Subsequently, a neural network (NN) router selectively identifies relevant documents from the document corpus. This sub-selection optimizes the accuracy and efficiency, reducing the number of documents loaded in the preliminary retrieval (step 3), which provides the first round of context used to further refine the queries (step 4). Following this, the retrieval stage utilizes the NN router to select the documents (step 1) on which the RAG realizes the second retrieval (step 2). Note that using the improved query boosts the accuracy of the second retrieval thanks to the more accurate embedding representations. The pipeline finalizes with a generating component, relying on a state-of-the-art language model such as GPT-3.5, which generates responses based on retrieved context.

\subsection{Hyperparameters Optimization}

% {\color{ali}[Ali]:No need to capitalize the words retrieval augmented. We try to avoid capitalizing inside the body of the paper unless its an acronym .}
As numerous studies have shown, hyperparameter optimization can provide large gains for retrieval-augmented models (see \cite{siriwardhana2023improving, llamaindex2024}). Therefore, using a synthetic dataset constructed for this purpose, we conducted a meticulous optimization of the chunk size, context length, indexing strategy, and embedding models (see Sec. \ref{section:experiments}). These hyperparameters are explained below 
% \textcolor{blue}{Antonio: never use the world model alone; explain which model are you talking about}:
\begin{itemize}
  \item \textbf{Chunk Size:} Determines the length of each text segment the RAG processes at once.
  \item \textbf{Context Length:} Length of the context yielded by retrieval component.
  \item \textbf{Embedding Models:} Algorithms that transform text into numerical representations.
  \item \textbf{Indexing Strategy:} The FAISS index\footnote{ FAISS is a library by Facebook AI Research optimized for efficient similarity search and clustering of dense vectors in large datasets on modern CPUs and GPUs.} \cite{johnson2017faiss} by which the model assesses the relevance of each text chunk related to the given query. 
  
\end{itemize}

 % {\color{done}[Ali]FAISS acronym definition as its not common knowledge}
 % \sout{This paper involved an evaluation methodology of advanced retrieval algorithms. Central to this evaluation was the deployment of} 
 % \textcolor{blue}{Antonio: mode this when you first mentione FAISS, as a footnote, while keeping the reference in the main text}

% \textcolor{green}{what does FAISS mean? what do you want to retrieve? references?}: IndexFlatIP, IndexFlatL2, and IndexHSNWFlat. 
\subsection{Query Augmentation}
\label{sec:Query Augmentation}
Numerous studies indicate significant improvements when augmenting vague queries in RAG pipelines \cite{chan2024rqrag}. In telecom documents, two major issues arise with vague queries: the abundance of technical terms and abbreviations in questions, and the inability of the RAG to discern user intent, leading to the retrieval of irrelevant, albeit similar, information.

\subsubsection{Lexicon-enhanced Queries}
% {\color{ali}[Ali]:Same issue Andrei, no need to redefine LLMs again. Just use the acronym.}

In this section, we address the challenge posed by the prevalence of technical terms and abbreviations in questions, which are often difficult to capture accurately in the embedding space. To tackle this issue, we utilized the ``Vocabulary for 3GPP Specifications" \cite{3GPPTR21.905} to construct two dictionaries: one for abbreviations and another for terms with their definitions. Integrating these dictionaries into our query enhancement block of the pipeline- see Fig. \ref{fig: pipeline}- allowed us to refine the embedding process. For each question, we enriched the embedding with the definitions of relevant terms from the dictionaries. This process refines the similarity evaluation between the question and potential answers by incorporating domain-specific knowledge. We also integrated relevant terms from the dictionaries in our final prompt, built using the retrieved context, user query, and the defined terms and definitions. This ensures that the LLM was prompted with the necessary technical vocabulary and definitions to process the question effectively. This method was employed in the Glossary Enhancement block of our pipeline, see Fig. \ref{fig: pipeline}.

% \textcolor{blue}{Antonio: what is the retrieval prompt? why final? this is the first time you mention it}
 % \textcolor{cyan}{[Nicola: I'm not sure 'equipped' is the right word. I would use 'prompted']}

\subsubsection{Generating Candidate Answers}
We use a language model to generate all plausible answers based on the preliminary context selected in Retrieval 1. Then, we add these generated candidate answers to enhance the user's query, clarifying its intent and preventing the retrieval of irrelevant information. The embedded enhanced query improves the identification of the relevant information in the corpora, yielding a superior final answer quality. 

% {\color{done}[Ali]: It is not clear what has been done here. Perhaps an example? A figure? What kind of refinement is done? The text is very general.}
% \textcolor{blue}{Antonio: Indeed the enhancement process is not clear. how do you refine them? the LLM is only used to generate candidate answers? how many of them? and how this candidate answers are used?}

\subsection{Enhancing the RAM Usage of the Telco-RAG}
% \textcolor{blue}{Antonio: I have the feeling you repeat the same concept twice. Also, we do not yet that smaller chunk sizes improve text retrieval accuracy. so make the relation between this and the router later}
\label{sec: RAG's RAM usage}
For large document corpora, the dataset of the embedded chunks becomes so voluminous that it exceeds the limitations of RAM capacities. Besides, we show in this work that for highly technical documents, smaller chunks yield better performance (see Sec. \ref{sec:Chunk Size optimization}). However, the smaller the chunks, the more the text segments to be processed by the RAG, which increases the required RAM resources. To deal with this issue, we recall that the 3GPP standards categorize specifications into 18 distinct series \cite{specsbyseries}. Each series provides the technical details of a specific aspect of mobile telecommunications technologies (radio access, core network components, security, etc). To improve the RAG usage efficiency, we developed an NN router tailored to predict relevant 3GPP series based on queries. This model enables selective loading of embeddings, thus drastically reducing the RAM usage.

% {\color{ali}[Ali]: Same problem with NN capitalization and redefinition.}\sout{To address the challenges posed by increased memory consumption due to smaller chunk sizes} \textcolor{blue}{To improve the RAG memory usage efficiency}, we developed a Neural Network (NN) model tailored to predict relevant 3GPP series based on queries. This model enables selective loading of embeddings, thus drastically reducing memory usage.

% As an example, Figure~\ref{fig: tsneplot} illustrates the separability between randomly selected documents from each 3GPP series suggesting that implementing an NN router can preserve if not enhance the RAG's accuracy.

 % \textcolor{blue}{Antonio: not clear}
% {\color{done}[Ali]:Improve this sentence}
% ({\color{done}[Ali]:Why 18?.})
%  {\color{ali}[Ali]:Perhaps a figure of the architecture could be helpful here? It is hard to understand the alpha and beta how they are incorporated.}.
%  {\color{done}[Ali]:Same comment here, it is hard to follow without a figure or some way to illustrate what is going on.}.
 % \textcolor{blue}{Antonio: what is a product feature?}
 
The architecture of the NN router incorporates two distinct input channels. The first channel processes {input 1,} a 1024-sized vector embedding the initial user query, while the second channel processes {input 2,} an 18-sized vector. Each entry of this vector is defined as the inner product between query embeddings and the embedding of each 3GPP series summary description, generated through a dedicated LLM.

% \textcolor{blue}{Antonio: why do you talk about description? what is this?}. %The utilization of 18 features strikes a balance between thorough analysis and computational manageability, optimizing the system’s complexity and efficiency. This configuration facilitates dynamic adaptations in relevance allocation across inputs during the determination of the appropriate 3GPP series.

\begin{figure}[t]
\centering
\includegraphics[width=0.48\textwidth]{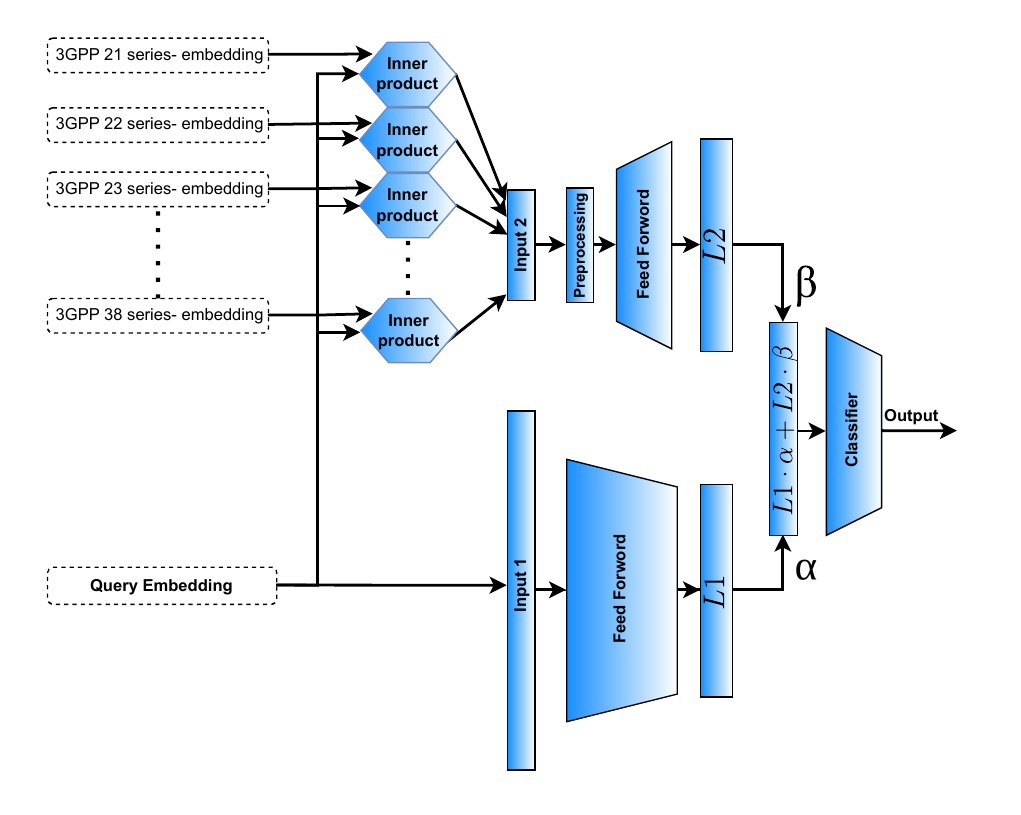}
\caption{The proposed NN router architecture.}
\label{fig: tsneplot}
\vspace{-1.8em}
\end{figure}

Central to our model are two adjustable trainable parameters, \(\alpha\) and \(\beta\), which modulate the influence exerted by each input stream on the resultant output. The overall architecture is illustrated in Fig.~\ref{fig: tsneplot}.

For the processing of the embedded query, our model implements a series of linear transformations that reduce its dimensionality from 1024 to 256. This reduction incorporates dropout layers to mitigate overfitting and a batch normalization layer to enhance training stability. Concurrently, the second input stream begins with a dimensionality of 18, {preprocessed through a softmax layer}, which is then expanded to 256 dimensions, to process jointly the contributions from both input streams in the decision-making process. The outputs from these pathways are weighted by 2 trainable parameters, \(\alpha\) and \(\beta\). These weighted outputs are summed up into a unified representation, which our neural network model utilizes to ascertain the target 3GPP series with heightened accuracy.

% \textcolor{blue}{Antonio: this looks like a conclusion. No need of the following paragraph here. Focus more on better explaining your work}
Integrating this NN model into Telco-RAG framework significantly elevates the ability to discern and categorize standards-related queries, paving the way for more targeted and efficient information retrieval. 

% {\color{ali}reduce the claim of last sentence or remove it for space reduction}
% \textcolor{blue}{Antonio: not clear why the following paragraph is here}
To train the NN router, we created a synthetic dataset comprising 30,000 questions from 500 documents from 3GPP Release 18, and their originating series that served as target labels. The adoption of synthetic data for training and testing our NN router reduces the risk of overfitting the dataset on which we test Telco-RAG pipeline \cite{Gilardi2023}.

\subsection{Prompt Engineering}\label{subsec:prompt}
Prompt engineering plays a crucial role in RAG systems, particularly in ensuring that the RAG maintains focus on the user's question while comprehending the broader context \cite{unleashing2023prompt}. 

% \textcolor{cyan}{[Nicola: To much fancy wording IMO. This does not help the reader to understand. Why not say directly to the reader that in this section we design the prompt, and that we want to provide the retrieved information in the most effective way to the LLM?]}
In our study, we designed a structured, dialogue-oriented prompt, as prompt engineering literature has shown better LLM performance with this format \cite{Gao2019NeuralAT}. More specifically, the final prompt of Telco-RAG starts with the query followed by the definitions of the terms and abbreviations. After that, the prompt includes the generated context. Importantly, the proposed prompt includes the query repetition, before the related options and query instruction helping the model to effectively generate a relevant response. The designed format of the LLM prompt is as follows:

% {\color{done}[Ali]:Same things with the words i mentioned before.}

{
\begin{tcolorbox}[enhanced,
                  colback=promptbackground, 
                  colframe=black,
                  boxrule=0.5pt,
                  arc=2mm,
                  top=2mm, bottom=2mm, left=1mm, right=1mm, 
                  boxsep=5pt, 
                  listing only,
                  listing options={ 
                      basicstyle=\tiny\ttfamily, % Changed from \small to \tiny
                      breaklines=true 
                  }]
                  
*Please provide the answers to the following multiple-choice question: $\textless$Question$\textgreater$\\
        *Terms and Definitions: $\textless$Defined Terms$\textgreater$\\
        *Abbreviations: $\textless$Abbreviations$\textgreater$\\
        *Considering the following context:
        $\textless$Retrieved Context$\textgreater$\\
        *Please provide the answers to the following multiple-choice question: $\textless$Question$\textgreater$\\
        *Options: $\textless$Options$\textgreater$ \\
        *Write only the option number corresponding to the correct answer.
        
        % {\color{done}[Ali]:Why we put the question many times?.}
\end{tcolorbox}
}

% {\color{done}[Ali]:Use smaller titles (here and before too).}

\begin{table*}[t]
\caption{Architectures of the compared RAGs.}
\centering
\small 
\renewcommand{\arraystretch}{1.2} 
\setlength{\tabcolsep}{3pt} 
\begin{tabular}{@{}>{\raggedright\arraybackslash}p{3cm} 
                  >{\raggedright\arraybackslash}p{2.5cm}
                  >{\raggedright\arraybackslash}p{1.2cm}
                  >{\raggedright\arraybackslash}p{2cm}  
                  >{\raggedright\arraybackslash}p{2cm}  
                  >{\raggedright\arraybackslash}p{2cm}  
                  >{\raggedright\arraybackslash}p{1.5cm}
                  >{\raggedright\arraybackslash}p{2cm}   
                  @{}}
\toprule
\textbf{Name} & \textbf{Embedding Model} & \textbf{Chunk Size} & \textbf{Context Length} & \textbf{Candidate Answers} & \textbf{Glossary Enhancement} & \textbf{NN Router} & \textbf{Enhanced Final Prompt} \\
\midrule
Benchmark RAG & -3-large[1024 size] & 500 & 1500-2500 & IndexFlatIP & No & No & No \\
Telco-RAG & -3-large[1024 size] & 125 & 1500-2500 & IndexFlatIP & Yes & Yes & Yes \\
\bottomrule
\end{tabular}
\label{tab: 2RAGS}
\vspace{-0.5em}
\end{table*}

\section{Experimental Results}\label{section:experiments}
In this section, we present the performance of Telco-RAG framework in enhancing the capabilities of LLMs applied to the telecom domain. To achieve this task, we have used two sets of multiple choice questions (MCQs), one optimization set and one evaluation set, specifically created to assess the knowledge of LLMs in telecommunications. The optimization set is composed of 2000 MCQs generated following the methodology presented in \cite{maatouk2023teleqna} and based on documents from 3GPP Rel.18. The second set consists of the 1840 TeleQnA MCQs related to 3GPP documentations \cite{maatouk2023teleqna}.
The purpose of Telco-RAG is to effectively help professionals with complex queries from telecom domain. The MCQ format, though very convenient for evaluation purposes, does not realistically correspond to the type of queries that will be submitted to the system. i.e., the user will likely do not provide any option to the LLM. Hence, we decided not to include the options in the retrieval process and use them solely to assess Telco-RAG accuracy. 
In the following results, accuracy measures the fraction of correct answers of Telco-RAG to the queries in the datasets.

Table~\ref{tab: 2RAGS} presents the main parameters of Telco-RAG and the RAG benchmark architecture compared throughout the following experiments.

%{TeleQnA -\sout{the first} \textcolor{blue}{a} benchmark dataset specifically created to evaluate the knowledge of LLMs in telecommunications \cite{maatouk2023teleqna}. \sout{This} \textcolor{blue}{TeleQnA is a} Multiple Choice Questions (MCQ) benchmark \sout{is designed to assess the performance of question-answering systems across} \textcolor{blue}{focusing on} a variety of telecommunications topics, including Lexicon, Standards Overview, and Standards Specifications. \sout{We will use TeleQnA to evaluate the final performance of our model.}}
%\textcolor{blue}{Antonio: we do not need to have this text both here and in the NN router section; I think here is its place so take the text from the previous section and make one short paragraph with the specific details. Be precise and avoid general, vague sentence. It is clear for what we use this dataset already with the first sentence, which foloows.}
%{To avoid overfitting the TeleQnA benchmark while optimizing our \textcolor{blue}{RAG} framework, we \textcolor{blue}{have} generated synthetic questions specifically focused on 3GPP standards from a concise set of documents. This methodology was essential in creating a controlled experimental environment, where we were able to compare the influence of each parameter effectively. It also enabled us to evaluate not only the accuracy of the final answer but also the precision of the retrieval by comparing the chunks selected by the pipeline with those used to generate the question. }

\subsection{Hyperparameters Optimization}

% \begin{mdframed}[backgroundcolor=gray!20, linewidth=0pt, innerleftmargin=10pt, innerrightmargin=10pt, innertopmargin=10pt, innerbottommargin=10pt]
% \noindent
% \begin{verbatim}
% Please provide the answers to the 
% following multiple choice questions. 
% The questions will be in a JSON format,
% the answers must also be in a
% JSON format as follows:
%  {
% "question 1": {
% "question": question,
% "answer":"option {answer id}:{answer
% string}"
% },
% ...
% }
% Each question will have a field called 
% 'context' that helps you find the answer 
% to the question
% Output:
% \end{verbatim}
% \end{mdframed}

% \begin{figure}[h!]
%     \includegraphics[width=\linewidth]{accuracywinrate.png}
%     \caption{Win rate comparison. Dark blue corresponds to  IndexFlatIP and light blue to IndexFlatL2}
%     \label{fig: Winrate}
% \end{figure}

% {\color{done}[Ali]:Smaller subsubtitles. Also discuss how you're going to evaluate everything (i.e., using teleqna specs. It is good to remind at the beginning of the experimental results section)}.

% \textcolor{cyan}{[Nicola: which model?]}

% {\color{ali}[Ali]:Use Smaller subsubtitles for the subsubsection too.}
\subsubsection{Selecting the Embedding Model}

%Our investigation rigorously compares the performance dynamics between two distinct embedding algorithms: text-embedding-3-large and text-embedding-ada-002. 
%\textcolor{blue}{Antonio: of which empirical evidence are you talking about? Provide a short introduction describing what are you comparing here and the main difference between the two models; describe the experiment before providing the results}
In this experiment, we compare the performance of two OpenAI embedding models for the Telco-RAG framework: 1) Text-embedding-3-large and  text-embedding-ada-002 \cite{openai2023embeddings}.
Text-embedding-3-large extends the capabilities of its predecessor text-embedding-ada-002. Text-embedding-3-large is trained using Matryoshka Representation Learning \cite{kusupati2022matryoshka}, a technique that allows the shortening of the embedding vectors, which reduces computational and RAM requirements, while preserving a stronger performance. Our results show that, on average, the text-embedding-3-large model, with a fixed embedding dimension of 1024, improves the accuracy of Telco-RAG by 2.29$\%$, over the text-embedding-ada-002 model.

  % \sout{\textcolor{orange}{ This indicates that the choice of the embedding model plays a critical role in the effectiveness of the retrieval phase and underscores the importance of selecting a better embedding model.}} \textcolor{blue}{Antonio: the following is too vague.} This result is not surprising as text-embedding-3-large is a newer and more robust model designed to better handle complex reasoning and language understanding tasks \textcolor{blue}{Antonio: how? state which model you use for the following experiments}. %MTEB benchmark results clearly illustrate that the text-embedding-3-large model is superior to the ada v2 model, with scores of 64.6 compared to 61.0, respectively. 

\subsubsection{Chunk Size Optimization}
\label{sec:Chunk Size optimization}
{We have assessed the influence of varying chunk sizes— 125, 250, and 500 tokens—on the accuracy of RAG systems. Importantly, there is an inverse relationship between chunk size and Telco-RAGaccuracy. 
%while RAM usage grows inversely proportional to chunk size. 
These results highlight the critical importance of optimizing chunk size, which has led to an average improvement of 2.9$\%$ in accuracy when selecting as chunk size 125 tokens instead of 500 tokens, for equal context length.}

\subsubsection{Context Length Optimization}
% \textcolor{blue}{Fig. \ref{fig: increasing_contextlength} shows the results of our experiments to optimize the context length in the proposed RAG framework.}
% \textcolor{blue}{Antonio: why do you talk here about chunk size optimization?}\textcolor{blue}{Our} findings indicated an inverse relationship between chunk size and system accuracy, leading to an inquiry into the optimal context length. 

\begin{figure}[t]
\centering
\includegraphics[width=0.9\linewidth]{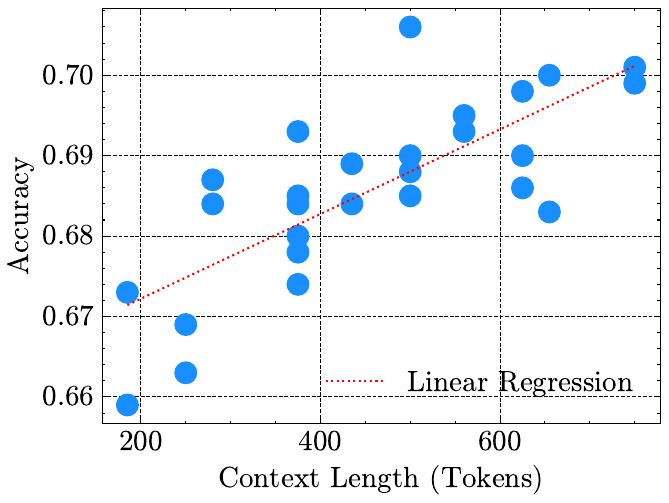}
\caption{RAG's accuracy vs context length. }
\label{fig: increasing_contextlength}
\vspace{-1.3em}
\end{figure}

% \textcolor{blue}{Antonio: how do you do this?}
Fig.~\ref{fig: increasing_contextlength} shows the linear regression fitted on the RAG accuracy computed for a diverse set of context lengths, with different configurations. The results show an ascending trend of the accuracy as a function of context length. As a side note, we have noticed a drop in performance when the context length gets larger than 1500 tokens. However, this is alleviated by presenting the query twice, before and after the context, as discussed in Sec. \ref{subsec:prompt}. 

% \textcolor{cyan}{[Nicola: The figure is not  introduced (just dropped in the middle of the text) and not well described.]}

\subsubsection{Indexing Strategy Selection}

In our research, we {have} evaluated the impact of different indexing strategies in the accuracy of Telco-RAG: 1) IndexFlatL2, 2) IndexFlatIP, and 3) IndexHNSW. 
IndexFlatL2 is based on the Euclidean distance while IndexFlatIP uses Euclidean dot product. In constrast, IndexHNSW is an approximate method for efficient searching in high-dimensional data spaces using Euclidean distance.
IndexHSNW has shown considerably inferior performance compared to IndexFlatIP and IndexFlatL2. Importantly, despite marginal differences in terms of accuracy, IndexFlatIP has outperformed IndexFlatL2 in 80$\%$ of our experiments.

% \begin{table*}[h]
% \centering
% \label{table:openai_comparison}
% \begin{tabular}{@{}llllll@{}}
% \toprule
% \textbf{Embedding model}  & \textbf{Query refinements}& \textbf{Chunk Size} & \textbf{Context length}  & \textbf{Accuracy} \\
% \midrule
% Text-embedding-3-large & Without refined query & 375 & 3000 & \textbf{0.7440} \\
% Text-embedding-ada-002 & Without refined query & 375 & 3000 & 0.7293 \\
% \midrule
% Text-embedding-3-large & Without refined query & 500 & 2000 &  \textbf{0.7386} \\
% Text-embedding-ada-002 & Without refined query & 500 & 2000 & 0.7158 \\
% \midrule
% Text-embedding-3-large & With refined query & 2000 & 8000 & \textbf{0.7875} \\
% Text-embedding-ada-002 & With refined query & 2000 & 8000 & 0.7745 \\
% \bottomrule
% \end{tabular}
% \vspace{0.5mm}
% \caption{Performance comparison between embedding models using the IndexFlatIP retrieval algorithm under various conditions.}
% \end{table*}

% {\color{ali}[Ali]:Use smaller subtites Andrei}

\subsection{Query Augmentation}
In this section we evaluate the gain in accuracy brought by enhancing the user queries through the methodology described in Sec. \ref{sec:Query Augmentation}.

\subsubsection{Lexicon-enhanced Queries}
To validate the effectiveness of this approach, we applied it to a subset of lexicon-focused questions from TeleQnA \cite{maatouk2023teleqna}, which were designed to evaluate the understanding of abbreviations and technical terms within the telecommunications sector. Our results presented in Table ~\ref{tab: lexicones} have shown that the designed RAG framework enhances the baseline LLM accuracy on lexicon questions, i.e., from 80.2 $\%$ to 84.8$\%$. However, Lexicon-enhanced queries have
achieved an accuracy rate exceeding 90$\%$ on these questions, gaining 6$\%$ compared to the same RAG pipeline without the lexicon enhancement.

\begin{table}[ht]
\centering
\caption{Impact of Lexicon-enhanced queries.}
\begin{tabular}{@{}lcc@{}}
\toprule
\textbf{Baseline (No Context)} & \textbf{Benchmark RAG} & \textbf{Telco-RAG} \\ 
\midrule
 80.2$\%$& 84.8$\%$ & 90.8$\%$ \\ 
\bottomrule
\end{tabular}
%\vspace{1mm}
\label{tab: lexicones}
\end{table}

\subsubsection{Enhancing User's Query With Candidate Answers}
To retrieve a better context, we enhance the user's query with candidate answers generated by an LLM (step 4 of the query's enhancement stage).
{Table \ref{tab:refined_query} presents the accuracy of Telco-RAG with and without the usage of these candidate answers. Specifically, we can observe that for the text-embed-ada-002 embedding model, the addition of candidate answers considerably improves the query embedding representations, bringing a 3.56$\%$ average accuracy gain. The accuracy of the RAG with text-embed-ada-002 including refined queries is larger than the one achieved using text-embed-3-large without refined queries. Furthermore, with text-embed-3-large, we observe a gain of 2.06$\%$ on average accuracy when using candidate answers in the retrieval process.}
 
 % \textcolor{cyan}{[Nicola: Table needs to be referenced explicitly: Table X shows blablabla ]}
\begin{table}[ht]
\caption{RAG's accuracy with and without refined query.}
\centering
\small % Reduce font size
\renewcommand{\arraystretch}{1.1} % Adjust spacing between rows
\setlength{\tabcolsep}{3pt} % Reduce spacing between columns
\begin{tabular}{@{}lcccc@{}}
\toprule
\textbf{Embedding} & \textbf{Chunk} & \textbf{Context} & \textbf{Initial} & \textbf{Refined} \\
\textbf{Model} & \textbf{Size} & \textbf{Length} & \textbf{Accuracy} & \textbf{Accuracy} \\ \midrule
Text-embed-ada-002 & 125 & 750 & 0.729 & \textbf{0.777 (+4.8\%)} \\
Text-embed-ada-002 & 250 & 2000 & 0.770 & \textbf{0.795 (+2.5\%)} \\
Text-embed-ada-002 & 500 & 2000 & 0.740 & \textbf{0.774 (+3.4\%)} \\
Text-embed-3-large & 125 & 750 & 0.744 & \textbf{0.780 (+3.6\%)} \\
Text-embed-3-large & 250 & 2000 & 0.784 & \textbf{0.796 (+1.2\%)} \\ 
Text-embed-3-large & 500 & 2000 & 0.774 & \textbf{0.788 (+1.4\%)} \\ \bottomrule
\end{tabular}
\label{tab:refined_query}
\end{table}

 % {\color{done}[Ali]: Similarly, words like remarkable, etc. need to be avoided.}
 % {\color{done}[Ali]:There is a lot of use of words such as strategically, rigorously, etc. When overdone, it seems that we are overselling what we are doing. I tried removing them but there are many of them so make sure you delete them. We should always try to avoid subjective adjectives and let our work speak for itself}
% {\color{done}[Ali]: No need to redefine acronyms.}

\begin{figure*}[t]
\centering
\includegraphics[width=0.6\linewidth]{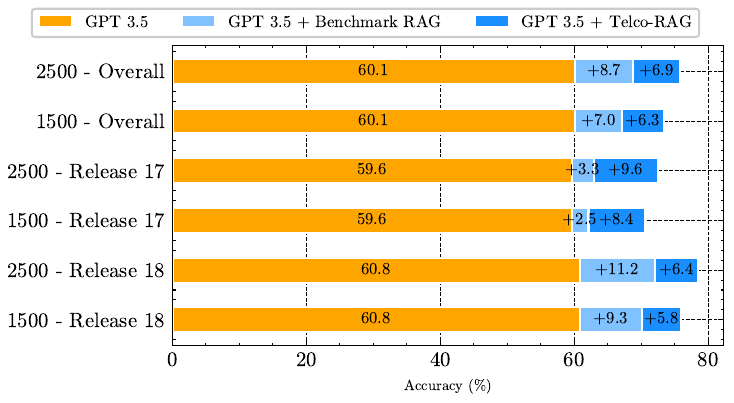}
\caption{Comparison the accuracy of Telco-RAG system with a baseline GPT 3.5 with/without the Benchmark RAG on the TeleQnA questions related to 3GPP documents.}
\label{fig: accimp}
\vspace{-1em}
\end{figure*}

\subsection{RAM Usage Analysis in the Telco-RAG}

\begin{figure}[ht]
\centering
\includegraphics[width=0.9\linewidth]{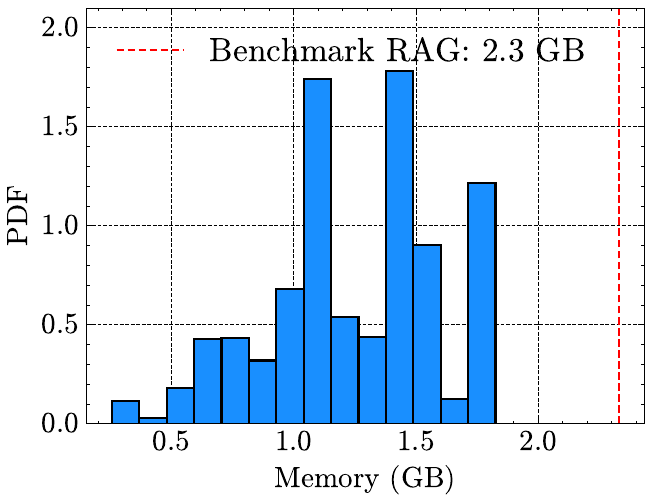}
\vspace{-1em}
\caption{PDF of the RAM usage of Telco-RAG vs Benchmark RAG.}
\label{fig: memory}
\vspace{-1em}
\end{figure}

% {\color{done}[Ali]:Remove remarkably}. 
Selecting a 125-token chunk size increases the RAM requirements of the Telco-RAG (see Sec. \ref{sec: RAG's RAM usage}). However, the integration of the designed NN router can tackle this issue.
% \textcolor{blue}{Antonio: let's present clearly Figure~\ref{fig: memory}. why there is a pdf? the reference is deterministic and the NN router makes the RAM requirement stochastic?}
{Fig.~\ref{fig: memory} presents the histogram of RAM usage for the 2000 MCQs in the optimization set. NN router dynamically selects the number of documents processed by the Telco-RAG pipeline based on their relevance to the query, as opposed to a fixed number of documents processed by the Benchmark RAG architecture. This method introduces variability in RAM usage among different queries, which results in the probability density function (PDF) in Fig.~\ref{fig: memory}. Our results show that the NN-enhanced RAG model leads in average to a RAM consumption of 1.25 GB, thus reducing of 45$\%$ the requirement of 2.3 GB -obtained by the Benchmark RAG solution.}

% \textcolor{blue}{Antonio: the usage of adverb in papers should be limited; especially in case like this. just say how gain the NN router brings}. 

As presented in Sec. \ref{sec: RAG's RAM usage}, the proposed NN router reduces the RAM usage of the designed RAG by selecting a limited number of retrieved documents, whose content relates with the query. To assess the capability of the NN router, we have compared it to GPT 3.5 and GPT 4 and provided to each model the questions of our optimization dataset including with each of their prompt the related 3GPP series description, Therefore, we have asked the three model to indicate the top k most related 3GPP series. Table \ref{tab: NN_compare} shows the accuracy determined by subsequently verifying if the correct 3GPP series was among the top k retrieved ones.

The results in Table~\ref{tab: NN_compare} highlight that the proposed NN model outperforms both GPT 3.5 and GPT 4 in identifying relevant 3GPP Series for a query, leading to an average accuracy gain of 37.8$\%$ and 11.1$\%$, respectively.

% \textcolor{blue}{Antonio: by how much?}
\begin{table}[ht]
\centering
\caption{Accuracy comparison of the NN router with GPT 3.5 and the GPT 4 at finding the most relevant 3GPP series for a given question. }
\begin{tabular}{@{}lcccc@{}}
\toprule
 \textbf{Top k} & \textbf{NN Router} & \textbf{GPT 3.5} & \textbf{GPT 4}\\ \midrule
 k=1 & \textbf{51.3}$\%$ & 19.9$\%$ & 30.4$\%$ \\ 
 k=3 & \textbf{80.6}$\%$ & 36.6$\%$  & 70.8$\%$\\
 k=5 & \textbf{88.3}$\%$ & 50.3$\%$  & 85.6$\%$\\ 
\bottomrule
%\vspace{0.5mm}
\end{tabular}
\label{tab: NN_compare}
\end{table}

The ability of the designed NN router to accurately deduce the applicable 3GPP series for a given query reduces the consideration of irrelevant content. This reduction not only lowers the computational complexity of the retrieval steps but also the overall resources needed for processing the retrieved content.

\subsection{Enhanced Prompt Formatting}

% \sout{The generation component's evaluation was structured to assess its precision and efficiency in handling MCQs. Moreover, we also assessed the model's adaptability to the quality and nature of the retrieved documents, influenced by the previously optimized chunk sizes and retrieval algorithms. In our experiments, we used advanced LLMs, such as GPT-3.5, for their next-generation capabilities.}  
% \sout{Our examination extends into the specialized analysis of the generation component, concentrating on the component's capacity to accurately select the correct answer from the provided options based on the information retrieved during the retrieval phase.}
In this section, we highlight the accuracy gain brought by the  prompt presented in Sec. \ref{subsec:prompt}, which we have designed for LLM answering MCQs related to telecom domain.
Our analysis of the results revealed a 4.6$\%$ average gain in accuracy, compared to the original JSON format of TeleQnA questions
% \sout{, for the same set of questions, a notable enhancement in the model's ability to parse and utilize context, underscoring the efficacy of this approach}.
This result suggests that human-like query structures can significantly elevate the contextual understanding and accuracy of LLM models.

\subsection{Overall Performance}
In this section, we present the accuracy of the Telco-RAG  on the evaluation MCQs, i.e., 1840 3GPP-related questions from TeleQnA \cite{maatouk2023teleqna}.
Specifically, we consider three groups of MCQs, Rel. 17 MCQs, Rel. 18 MCQs, and the overall set of TeleQnA MCQs related to 3GPP documentations. For each of these sets of MCQs, we compare the performance of GPT 3.5 with Telco-RAG, GPT 3.5 with the Benchmark RAG, and GPT 3.5 without RAG.
Fig.~\ref{fig: accimp} highlights that Telco-RAG leads to notable gains in all the experiments. Importantly, Telco-RAG results an average improvement of 6.6$\%$ and 14.45$\%$ compared to GPT 3.5 with and without the Benchmark RAG.

\section{Conclusions}
% {\color{ali}The conclusion is too long, aim to reduce it}
{This paper presented Telco-RAG, a novel RAG framework for processing 3GPP telecommunications standards and supporting LLM in telecom use cases. We have demonstrated that refinements in chunk sizes, embedding models, indexing strategies, and query structuring significantly boost RAG system performance and accuracy. The provided solutions are general and can deal with frequent challenges encountered in building RAG pipelines for highly technical domains. We expect that the Telco-RAG, which we make publicly available, and the associated results will contribute substantially to the integration of AI in the telecommunications field.}

% {\color{done}[Ali]: Make transitions between each point smoother as they sound bulky and one after another without transition.}
% {\color{done}[Ali]: Don't use abbreviations such as We've; always use we have (full form).}

\begin{acronym}[AAAAAAAAA]
 \acro{3GPP}{third generation partnership project}
 \acro{AI}{artificial intelligence}
 \acro{API}{application programming interface}
 \acro{BS}{base station}
 \acro{BERT}{bidirectional encoder representations from transformers}
 \acro{FPGA}{field-programmable gate array}
 \acro{GPT}{generative pre-trained transformer}
 \acro{IEEE}{institute of electrical and electronics engineers}
 \acro{ITU}{International Telecommunication Union}
 \acro{GPU}{graphical processing unit}
 \acro{LLM}{large language mod}
 \acro{MAPE}{mean absolute percentage error}
 \acro{MCQ}{multiple-choice questions}
 \acro{RAG}{retrieval-augmented generation}
 \acro{RAN}{radio access network}
 \acro{RLHF}{reinforcement learning with human feedback}
 \acro{SLM}{small language model}
 \end{acronym}

\bibliographystyle{IEEEtran}
\bibliography{reference.bib}

\end{document}